\documentclass{PoS}
\usepackage{slashed}
\usepackage{graphicx}
\usepackage{wrapfig}
\usepackage{amsmath,lineno}
\newcommand {\op}{\ensuremath{\Sigma}}

\newcommand{\hmu}{\hat{\mu}}
\def\lsim{\raise0.3ex\hbox{$<$\kern-0.75em\raise-1.1ex\hbox{$\sim$}}}
\def\gsim{\raise0.3ex\hbox{$>$\kern-0.75em\raise-1.1ex\hbox{$\sim$}}}

\title{Critical behavior and net-charge fluctuations from
	lattice QCD\footnote{Based on talks given at the conference "{\it The Critical
	Point and Onset of Deconfinement}" (CPOD 2018), Sept. 24-28, 2018, Corfu,
Greece; the EMMI workshop "{\it Probing the Phase Structure of Strongly 
Interacting Matter: Theory and Experiment}", March 25-29, 2019, GSI Darmstadt, 
Germany, and the EMMI Rapid Task Force meeting "{\it Dynamics of critical 
fluctuations: theory-phenomenology-HIC}", April 8-12, 2019, GSI Darmstadt, 
Germany.}}

\ShortTitle{Critical behavior and net-charge fluctuations from lattice QCD}

\author{Frithjof Karsch\\
	Fakult\"at f\"ur Physik, Universit\"at Bielefeld, D-33615 Bielefeld,
Germany\\
	Physics Department, Brookhaven National Laboratory, Upton, NY 11973, USA\\
	E-mail: \email{karsch@physik.uni-bielefeld.de}}

\abstract{
We present recent results on the critical and pseudo-critical temperatures in 
$(2+1)$-flavor QCD with a physical strange quark mass and two degenerate
light quark masses extrapolated to the chiral limit and tuned to the
physical value, respectively. We furthermore discuss implication of the 
observed low chiral phase transition temperature, $T_c^0=132_{-6}^{+3}$~MeV,
for the structure of cumulants of conserved charge fluctuations at 
vanishing baryon chemical potential and consequences for the possible location 
of the QCD critical endpoint in the QCD phase diagram at non-zero baryon 
chemical potential.
}

\FullConference{Corfu Summer Institute 2018 "School and Workshops on Elementary Particle Physics and Gravity"\\
	(CORFU2018)\\
	31 August - 28 September, 2018\\
	Corfu, Greece}

\begin{document}
\section{Introduction}

Understanding the phase structure of strongly interacting matter is 
one of the central goals in studies of the properties of strong
interaction matter at finite temperature and density
through large-scale numerical calculations in the framework
of lattice regularized Quantum Chromo Dynamics (QCD). Also experimentally
major efforts at the Large Hadron Collider (LHC) at
CERN and the Relativistic Heavy Ion Collider (RHIC) at Brookhaven
National Laboratory are devoted to this goal. 

\begin{wrapfigure}{l}{0.5\textwidth}
        \includegraphics[width=0.45\textwidth]{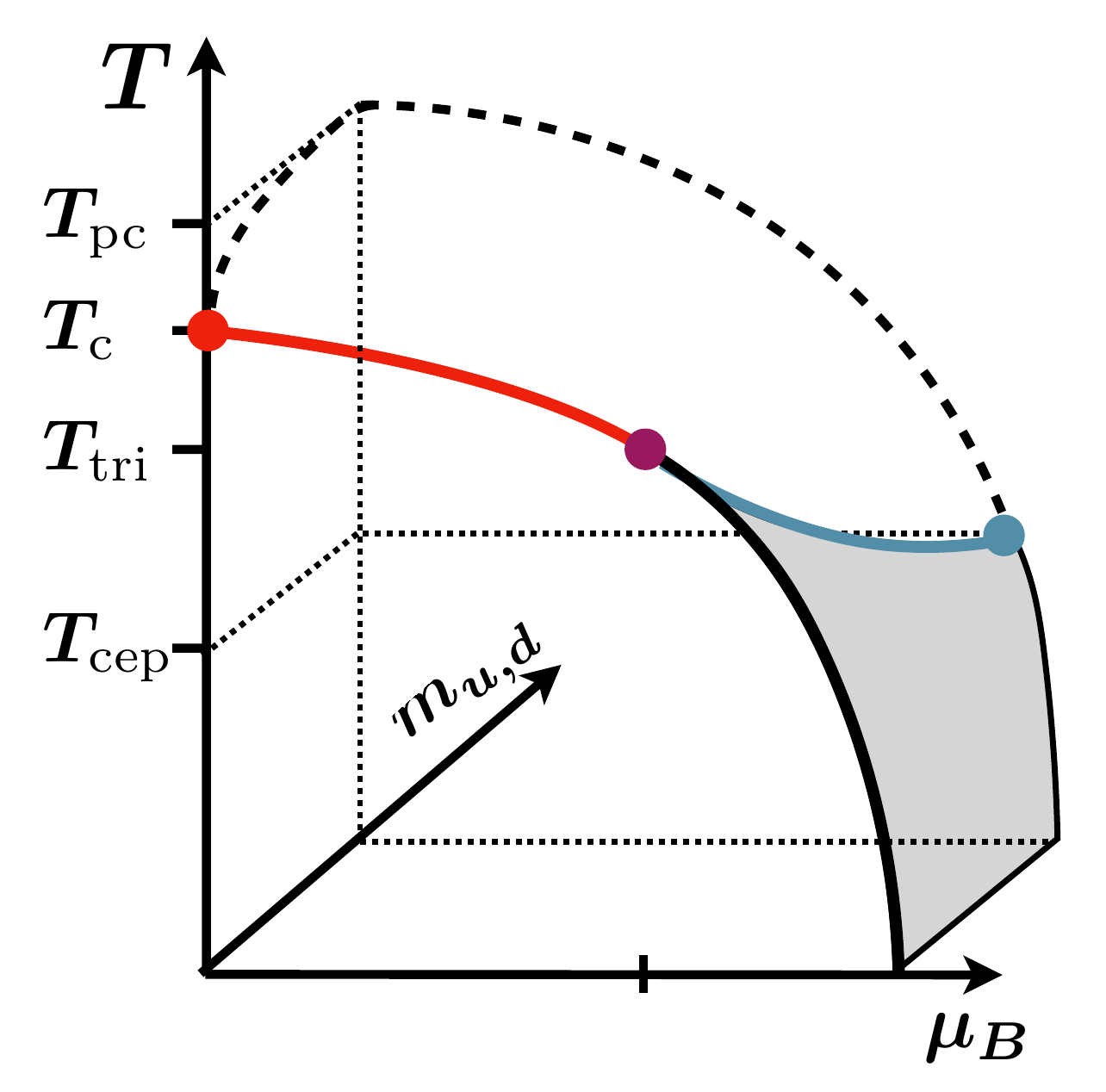}
        \caption{Sketch of a possible QCD phase diagram in the
		space of temperature ($T$), baryon chemical potential
		($\mu_B$) and light quark masses ($m_{u,d}$).}
\label{fig:phasediagram}
\end{wrapfigure}
\noindent
At vanishing net baryon-number density or,
equivalently, vanishing baryon chemical potential ($\mu_B$),
it is by now well established that the transition from
hadronic matter at low temperature to the quark-gluon plasma at
high temperature is a continuous (crossover) transition taking place
at a pseudo-critical temperature $T_{pc}$
(for recent reviews see \cite{Ding:2015ona,DElia:2018fjp}).
While this is the case for physical values of the quark masses,
it is expected that in the limit of vanishing light quark masses ($m_{u,d}$)
strong interaction matter shows true critical behavior
resulting from the appearance of second order phase transitions
at some temperature $T_c(\mu_B)$. 
In QCD with two massless quark flavors this transition is due to
the spontaneous breaking of the $SU_L(2)\times SU_R(2)\simeq O(4)$ chiral 
symmetry
\cite{Pisarski:1983ms} and persists as such also at non-zero baryon chemical 
potential. 

At non-zero values of the two light quark masses the transition 
is only a smooth crossover for small values of $\mu_B$. At larger 
$\mu_B$, however, it is expected that a second order phase transition arises
at the endpoint ($T_{cep}$) of a line of first order transitions, at which 
the net baryon-number density changes discontinuously \cite{Halasz:1998qr}. 
Critical behavior in the vicinity of this endpoint will be controlled
by the $3$-$d$, $Z(2)$ universality class. This Ising-like transition
will exist for arbitrary values of the light quark masses and thus
will meet the $O(4)$ chiral transition line at $m_{u,d}=0$ in a tri-critical
point ($T_{tri}$). A sketch of the resulting phase diagram, which also 
indicates the relative ordering of the various transition temperatures, 
is shown in Fig.~\ref{fig:phasediagram}. This generic phase diagram,
in particular the indicated ordering of the various characteristic
(phase) transition temperatures, is in
qualitative agreement with various model calculations
\cite{Halasz:1998qr,Stephanov:2006dn,Buballa:2018hux}.

In the following we will present recent lattice QCD results on the 
pseudo-critical ($T_{pc}$) and critical temperature ($T_c$) in 
$(2+1)$-flavor QCD at $\mu_B=0$. We relate these findings to 
the structure of higher order cumulants of conserved charge
fluctuations, and discuss how they constrain the location of a 
possible critical point at $\mu_B>0$ and physical values of the quark masses. 

\vspace{-0.2cm}
\section{Universal pseudo-critical and critical behavior}

\subsection{Pseudo-critical temperature in (2+1)-flavor QCD}

In the limit of vanishing up and down quark masses QCD possesses an exact
global symmetry, the chiral $SU_L(2)\times SU_R(2)$ flavor symmetry. This 
symmetry is spontaneously broken at low temperature, signaled by a 
non-vanishing chiral condensate ($\langle \bar\psi \psi \rangle$). 
Chiral symmetry is explicitly broken
due to the non-vanishing light quark masses. Nonetheless, this explicit
breaking is small enough for chiral symmetry providing a good, approximate
order parameter at non-zero temperature -- the chiral condensate
$\langle \bar{\psi} \psi\rangle$. Its variation with quark mass as well as 
temperature is large in a small temperature interval, which
leads to well defined peaks in the corresponding chiral ($\chi^\Sigma$) and 
mixed ($\chi_t$) susceptibilities. These maxima in the susceptibilities are
used to define pseudo-critical temperatures, which, in the limit of vanishing
quark masses, converge to the uniquely defined critical temperature for the
chiral phase transition.

  \begin{figure}[t]
        \centering
        \includegraphics[height=4.9cm]{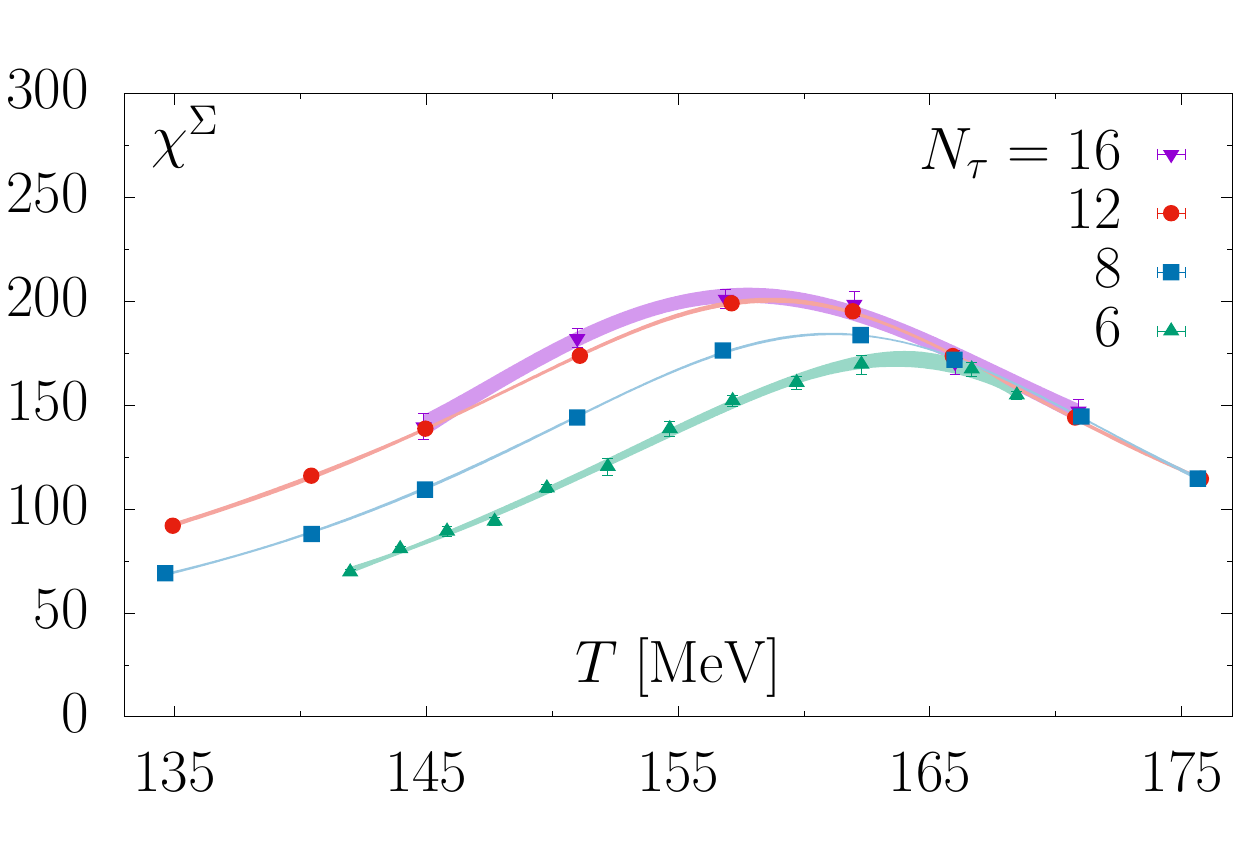}\hspace{0.4cm}
        \includegraphics[height=4.9cm]{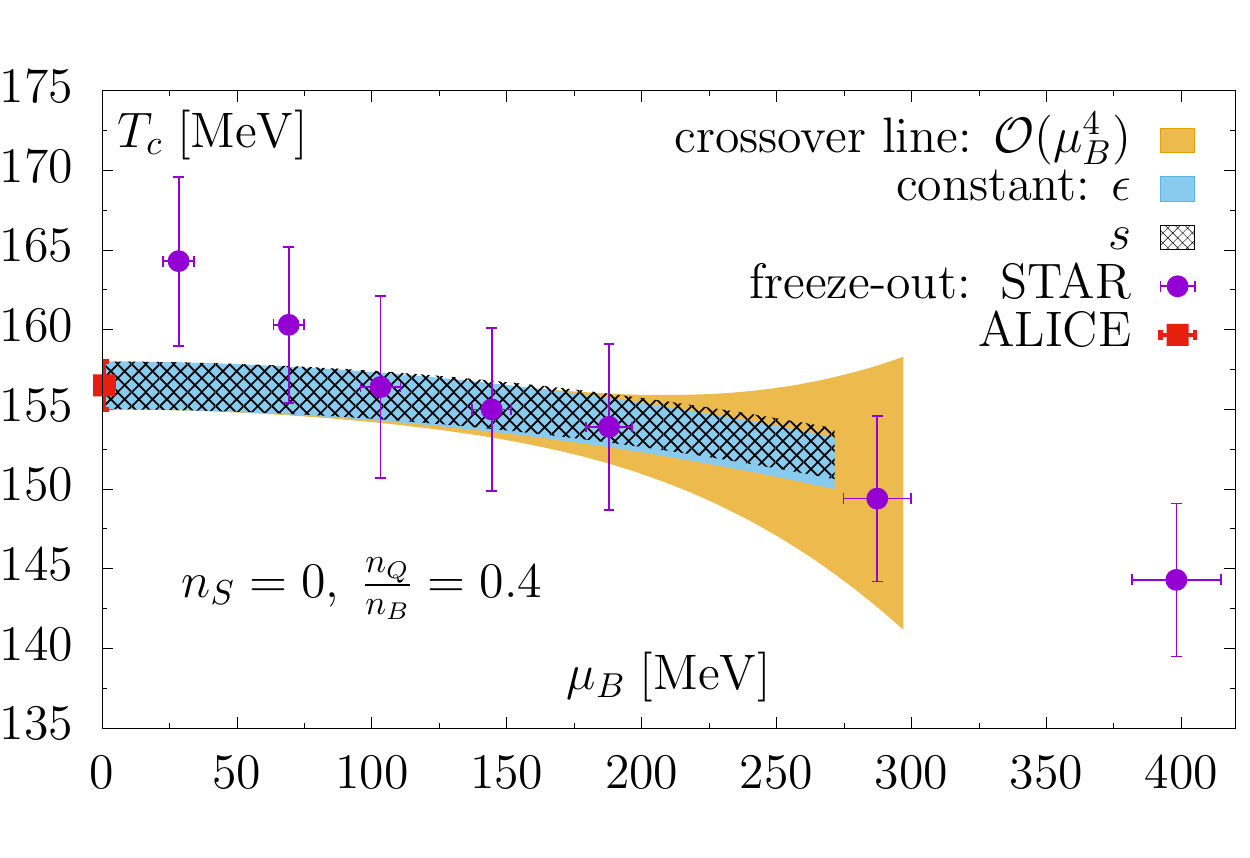}
        \caption{{\it Left:} The chiral susceptibility 
	($\chi^\Sigma\equiv \chi_M$) calculated on
        lattices with different temporal extent $N_\tau$ for physical
        values of the degenerate light $(u,d)$ and strange quark masses.
        {\it Right:} Crossover temperature $T_{pc}(\mu_B)$ determined
        from continuum extrapolated results for the location of peaks
        in the chiral susceptibilities defined in Eq.~\ref{eq:susg}
        and some further observables introduced in Ref.~\cite{Bazavov:2018mes}.
	Also shown in this figure are lines of constant energy and entropy 
	density \cite{Bazavov:2017dus} as well as results for freeze-out 
	temperatures determined from 
	data on particle yields measured by the STAR and ALICE collaborations
	\cite{Andronic:2017pug, Adamczyk:2017iwn}.
}       
\label{fig:Tpc}
\end{figure}

For our studies of the chiral phase transition we use as an order parameter
for chiral symmetry breaking 
\begin{equation}\label{eq:pbp}
  \op = \frac{1}{f_K^4} \left[
  m_s \left( \langle \bar\psi \psi \rangle_{u} + 
  \langle \bar\psi \psi \rangle_{d} \right)
  - (m_u+m_d)  \langle \bar\psi \psi \rangle_{s} \right] \, ,
\end{equation}
where $\langle \bar\psi \psi \rangle_f=T(\partial{\ln Z}/\partial{m_f})/V$ 
denotes
chiral condensates of the up (\(u\)), down (\(d\)), and strange (\(s\)) quarks.
A fraction of the strange quark chiral condensates is subtracted from the
light quark chiral condensates in order to eliminate ultra-violet divergences,
linear in the quark masses, and the condensates are multiplied with the
strange quark mass in order to define a renormalization group invariant
observable. The kaon decay constant $f_K$ is used to set the scale and define
a dimensionless order parameter $\Sigma$ (sometimes also denoted as $M$).

Pseudo-critical temperatures are extracted from the location of peaks in
the chiral and mixed susceptibilities
\begin{eqnarray}  \label{eq:susg}
   \chi_M &=&   m_s\left( \frac{\partial}{\partial m_u} +
   \frac{\partial}{\partial m_d} \right) \op \; , \label{chiM}\\ 
   \chi_t &=& T \frac{\rm d}{{\rm d}T} \op
   \,.
   \label{chit}
  \end{eqnarray}
For different values of the lattice spacing, $a=1/TN_\tau$, the peak locations
in different susceptibilities are determined. From an extrapolation
to the continuum limit, that takes into account ${\cal O}(a^2)$ cut-off
effects one then determines pseudo-critical temperatures for the
chiral transition.
Results from a recent determination of pseudo-critical temperatures
at physical values of the light and strange quark masses are shown in
Fig.~\ref{fig:Tpc}. The left hand figure shows the chiral susceptibility
($\chi^\Sigma\equiv \chi_M$) calculated on different size lattices 
($N_\sigma^3N_\tau$, with $N_\sigma = 4N_\tau$) \cite{Bazavov:2018mes} using
the Highly Improved Staggered Quark (HISQ) action \cite{Follana:2006rc}. Other
observables, e.g. the mixed susceptibility $\chi_t$, yield pseudo-critical 
temperatures, which in the continuum limit differ from each other by less 
than $2$~MeV \cite{Bazavov:2018mes}.
For the pseudo-critical temperature this analysis yields,
\begin{equation}
	T_{pc} = (156.5\pm 1.5)~{\rm MeV} \; .
	\label{Tpc}
\end{equation}
A comparison of this pseudo-critical temperature with the freeze-out 
temperature determined from data on particle yields in heavy ion
collisions at the LHC \cite{Andronic:2017pug} suggests that the formation of 
hadrons after the cooling of the expanding hot and dense quark-gluon matter 
created 
in these collisions does take place close to the phase boundary characterized
by this pseudo-critical temperature (see Fig.~\ref{fig:Tpc}~(right)).

\begin{figure}[t]
        \centering
        \includegraphics[height=4.9cm]{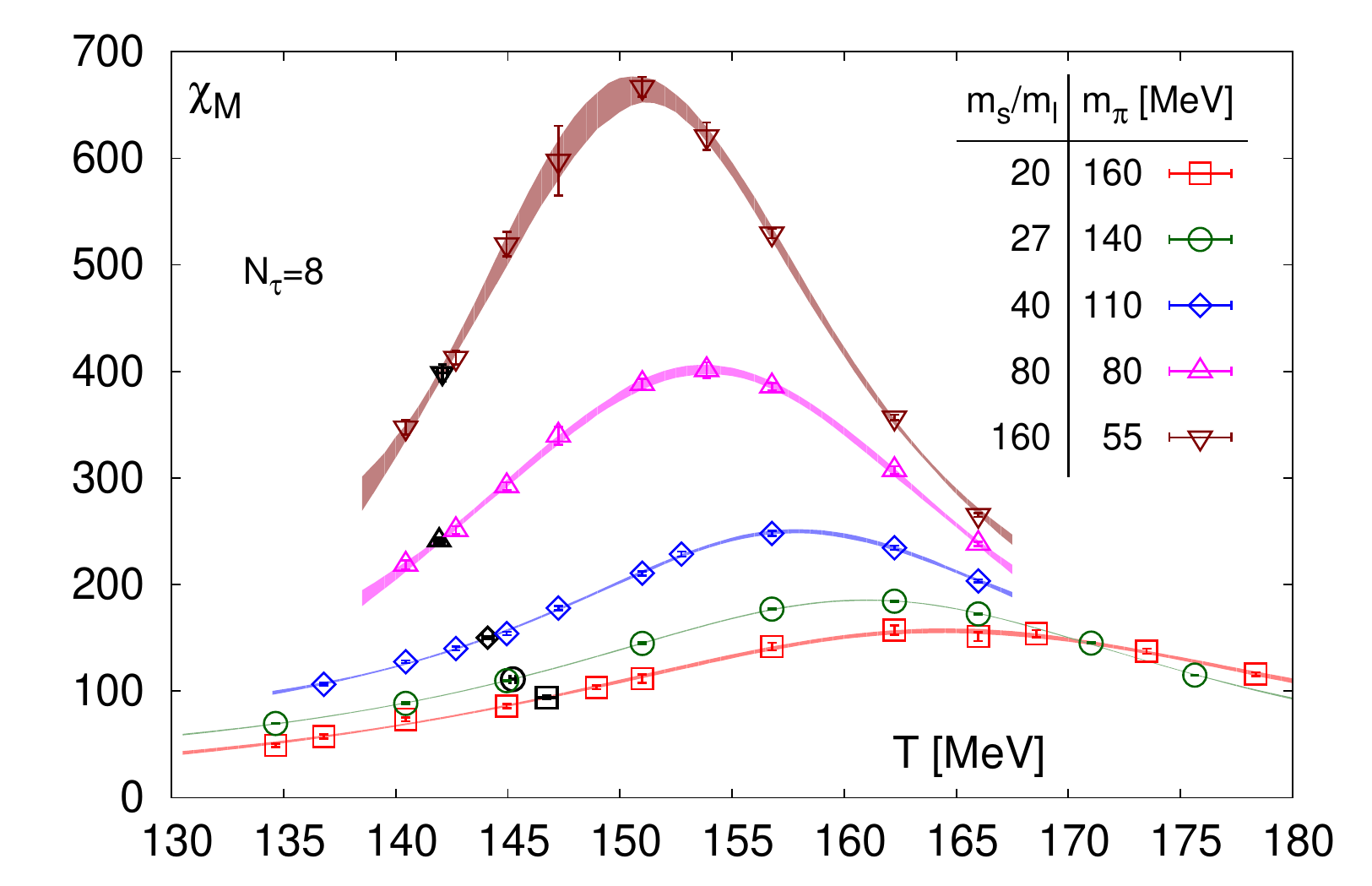}\hspace{0.4cm}
        \includegraphics[height=4.9cm]{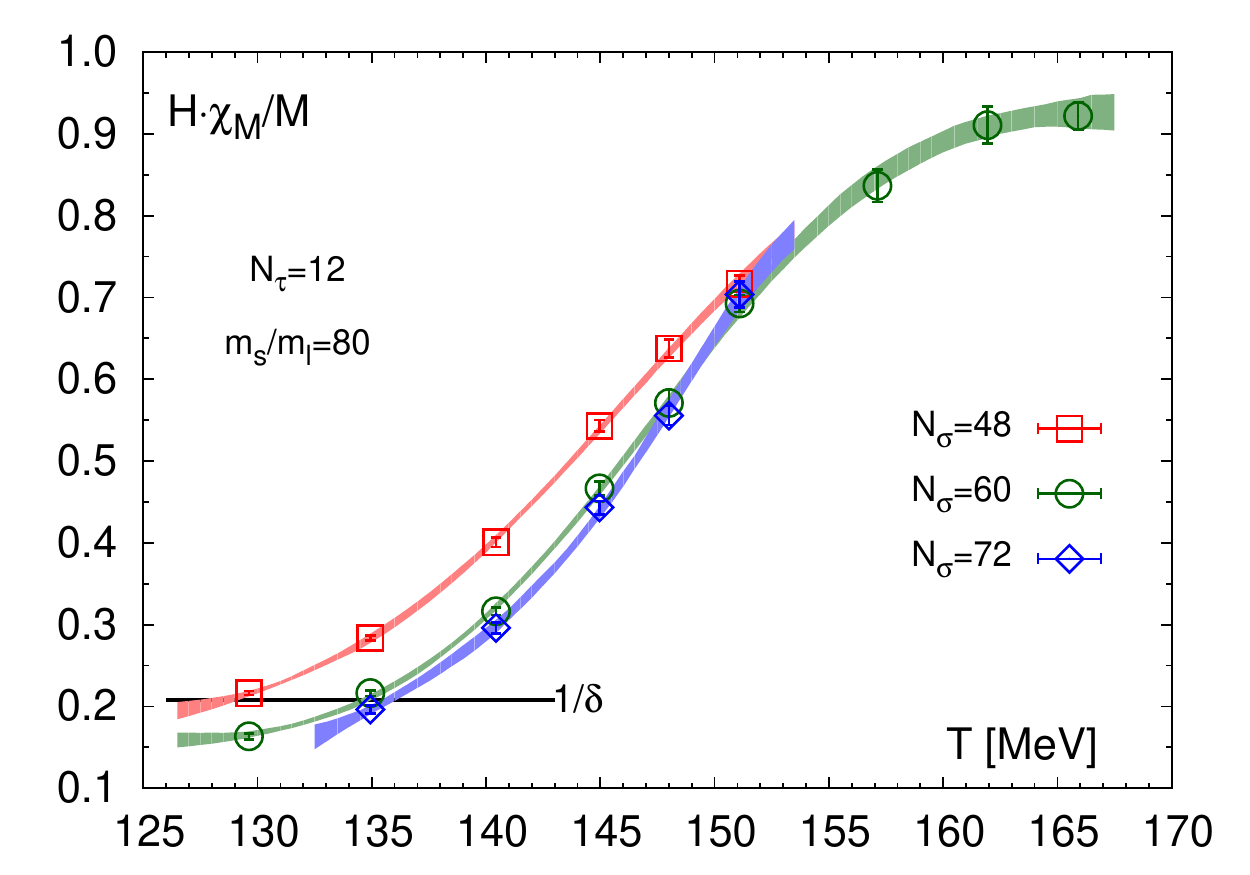}
        \caption{{\it Left:} The chiral susceptibility for several values
        of the quark mass ratio $H=m_l/m_s$ on lattices with temporal extent
        $N_\tau=8$ and spatial lattice sizes that are varied in the range 
	$N_\sigma = (4-7)$ when going from the largest to the smallest light 
	quark mass value.
	{\it Right:} The ratio $H \chi_M/M$ for $H=1/80$ and $N_\tau=12$
	for three different spatial lattice sizes $N_\sigma$.
}
\label{fig:chiralTc}
\end{figure}
\subsection{Critical temperature in (2+1)-flavor QCD}

An analogous analysis can be performed for other values of the light
quark masses ($m_l\equiv (m_u+m_d)/2$), keeping the strange quark mass fixed 
at its physical
value. The approach to the chiral limit, $H\equiv m_l/m_s\rightarrow 0$,
can then be examined by monitoring the quark mass dependence of the chiral 
order parameter and its susceptibility ($\chi_M$). Some results for the quark 
mass dependence of $\chi_M$, calculated with the HISQ action, are shown in 
Fig.~\ref{fig:chiralTc}~(left)
\cite{Ding:2019prx}. For sufficiently small values of the light quark masses
and close to the chiral transition temperature, {\it i.e.} in the scaling 
regime, the peak location in $\chi_M$, and 
similarly in $\chi_t$, is controlled by universal scaling functions,
\begin{equation}
	\chi_M(T,H) \sim h^{1/\delta-1}f_\chi(z)~+~regular \;\; , \;\; 
	\chi_t(T,H) \sim h^{1/\delta-1/\beta\delta}f'_G(z)~+~regular\; ,
	\label{scaling}
\end{equation}
where $\beta$ and $\delta$ are critical exponents for the
universality class of the chiral transition,
$z\equiv z_0[(T-T_c^0)/T_c^0]/H^{1/\beta\delta}$, $h=H/h_0$ and
$h_0$, $z_0$ are non-universal constants. The 
peak locations in $\chi_M$ and $\chi_t$ are related to maxima of the
scaling functions $f_\chi(z)$ and $f'_G(z)$, respectively. 
The quark mass dependence
of pseudo-critical temperatures thus is controlled by the scaling
variable $z$. The increase of the peaks is controlled by the prefactors. 
As can be seen in Fig.~\ref{fig:chiralTc}~(left) the peak in $\chi_M$
increases rapidly with decreasing quark, or 
equivalently pion, mass and the peak location shifts towards smaller 
values of the temperature. In the scaling regime, close to the chiral limit, 
contributions from regular terms will be small and one expects to find
\begin{equation}
	T_{pc} (H) = T_c^0 \left( 1 + \frac{z_X}{z_0} H^{1/\beta\delta}\right) \; ,
	\label{Tpc}
\end{equation}
with $z_X$ being a universal constant defining the location of the maximum
in $\chi_X$, e.g. $X\equiv M$ or $t$
when using the peak locations of $\chi_M$ and $\chi_t$ defined in 
Eq.~\ref{chiM} and Eq.~\ref{chit}, respectively. For the $3$-$d$, $O(4)$
universality class one has, $z_M\simeq 1.4(1)$, $z_t\simeq 0.8(2)$, and
$1/\beta\delta \simeq 0.55$
\cite{Engels:2011km}.
As $z_0$ typically is of ${\cal O}(1)$,
Eq.~\ref{Tpc} suggests that the pseudo-critical temperatures determined
from the peak locations in $\chi_M$ and $\chi_t$ will show a rather strong
dependence on the light quark masses. 
In fact, QCD-inspired model
calculations~\cite{Berges:1997eu,Braun:2005fj} suggest that $T_c^0$ might be 
$(20-30)$~MeV smaller than $T_{pc}$ calculated for physical values of the 
quark masses, for which $H\simeq 1/27$.

In order to determine the chiral 
phase transition temperature $T_c^0$ it thus would be advantageous to use 
observables which similarly to the maxima in susceptibilities correspond to 
a fixed value of the scaling variable $z$, but are related to a value
$z\equiv z_X$ that is close to zero. Two such
observables have been utilized recently \cite{Ding:2019prx} for this
purpose. One may define two characteristic temperatures, $T_\delta$ and
$T_{60}$, through the relations
\begin{eqnarray}
	\frac{H\chi_M(T_\delta)}{M(T_\delta)} &=&\frac{1}{\delta}
	\;\; , \label{Tdelta}\\
	\chi_M(T_{60}) &=& 0.6 \chi_M^{peak} \; .
	\label{T60}
\end{eqnarray}
In the thermodynamic limit the corresponding scaling variables 
$z_\delta$ and $z_{60}$ both are close to zero.
The resulting
estimators, $T_\delta$ and $T_{60}$, for the chiral phase transition
temperature are quark mass dependent only due to the presence of
contributions arising from regular terms in the partition function.
They therefore provide good estimators for the chiral phase 
transition temperature.
Some results for the ratio $H\chi_M /M$, from which the estimator
$T_\delta$ is extracted, are shown in Fig.~\ref{fig:chiralTc}~(right).
When decreasing the quark masses towards the chiral limit finite volume 
effects increase and some care needs to be taken in the extrapolation
to the thermodynamic limit.
After (i) infinite volume,  (ii) continuum, and (iii) chiral limit
extrapolations these estimators yield for the chiral phase transition
temperature \cite{Ding:2019prx}
\begin{equation}
	T_c^0 = 132^{+3}_{-6}~{\rm MeV} \; .
	\label{Tc0}
\end{equation}
The chiral phase transition temperature thus is about 25~MeV smaller
than the pseudo-critical temperature extracted from the location of 
the peak in the chiral susceptibility. As will be discussed further 
in Section~3, this has consequences also for the
phase transition temperature $T_{cep}$ at which a possible critical point
at physical values of the light quark masses and at 
non-zero values of the baryon chemical potential may occur.

\subsection{Curvature of the phase transition line in the chiral limit}

Close to the chiral limit, in the vicinity of the critical temperature,
the non-analytic (singular) behavior of the logarithm of the partition
function, {\it i.e.} the pressure, is described by a scaling function, $f_s(z)$.
Deviations from scaling are given in terms of an analytic (regular)
function $f_r$, 

\begin{equation}
	\frac{P}{T^4} = h^{2-\alpha} f_s(z) + f_r(T,\mu_B,\mu_Q,\mu_S,m_f) \; , 
\label{pressure}
\end{equation}
The reduced temperature variable $t$ entering the scaling variable 
$z\sim t/h^{1/\beta\delta}$ will also
dependent on the chemical potentials. In leading order,
and for vanishing strangeness and electric charge chemical potentials,
one has
\begin{equation}
	t\sim \frac{T-T_c^0}{T_c^0} + \kappa_2^{B,0} \left( \frac{\mu_B}{T} \right)^2
        \; ,
        \label{reduced}
\end{equation}
which also reflects the temperature dependence of the chiral phase
transition temperature, $T_c(\mu_B)= T_c^0 (1-\kappa_2^{B,0} ( \mu_B/T )^2)$.

At physical values of the quark masses the curvature of the transition
line, $\kappa_2^B$, will in general differ from $\kappa_2^{B,0}$, receiving
corrections from regular terms, terms arising from universal 
corrections-to-scaling or higher order terms in the scaling
variables being proportional to $H (T-T_c^0)$. This curvature term
can be extracted from the $\mu_B$-dependence of the location of maxima of
various susceptibilities. Using a Taylor expansion of, e.g. the mixed chiral
susceptibility $\chi_t(T,\mu_B)$ in terms of temperature and
baryon chemical potential around the pseudo-critical point $(T_{pc},\mu_B=0)$,
one obtains for the curvature $\kappa_2^{B}$ \cite{Bazavov:2018mes},
\begin{equation}
        \kappa_2^{B} = \left. \frac{1}{ 2 T^2 \partial_T^2 \chi_t }
       \left[ T \partial_T \chi'_t - 2 \chi'_t \right] \right|_{(T_{pc},\mu_B=0)} \; ,
       \label{kappat}
\end{equation}
with
$\chi'_t = T^2 \partial^2 \chi_t/\partial \mu_B^2$. Similarly one can
derive expressions for higher order expansion coefficients of $T_{pc}(\mu_B)$.
The analysis performed in Ref.~\cite{Bazavov:2018mes}
gave $\kappa_2^B=0.015(4)$ in agreement with other recent determinations
of the leading order correction to $T_{pc}$ 
\cite{Bellwied:2015rza,Bonati:2018nut}. 
The next-to-leading order correction,
$\kappa_4^B$, is an order of magnitude smaller and consistent with zero
within current statistical errors. The resulting $\mu_B$-dependence of the
crossover line for physical quark masses is shown in Fig.~\ref{fig:Tpc}~(right).

In the limit of vanishing quark mass
the curvature coefficients $\kappa_2^B$ will approach the corresponding
curvature term of the chiral phase transition line, $\kappa_2^{B,0}$.
In fact, in the absence of contributions
from regular terms the curvature coefficient will be quark mass
independent, as seen from the general scaling ansatz given in
Eq.~\ref{pressure}. To what extent this holds true may be probed by
comparing temperature and chemical potential derivatives of $P/T^4$.
In the absence of substantial contributions from regular terms
one expects to find in the scaling regime,
\begin{equation}
         \frac{T^2}{2} 
         \frac{\partial^2 \Sigma}{\partial \mu_B^2}=
        \kappa_2^{B,0} T_c^0 \frac{\partial \Sigma}{\partial T} 
        \; .
	\label{sym}
\end{equation}

\begin{wrapfigure}{l}{0.5\textwidth}
        \includegraphics[height=4.5cm]{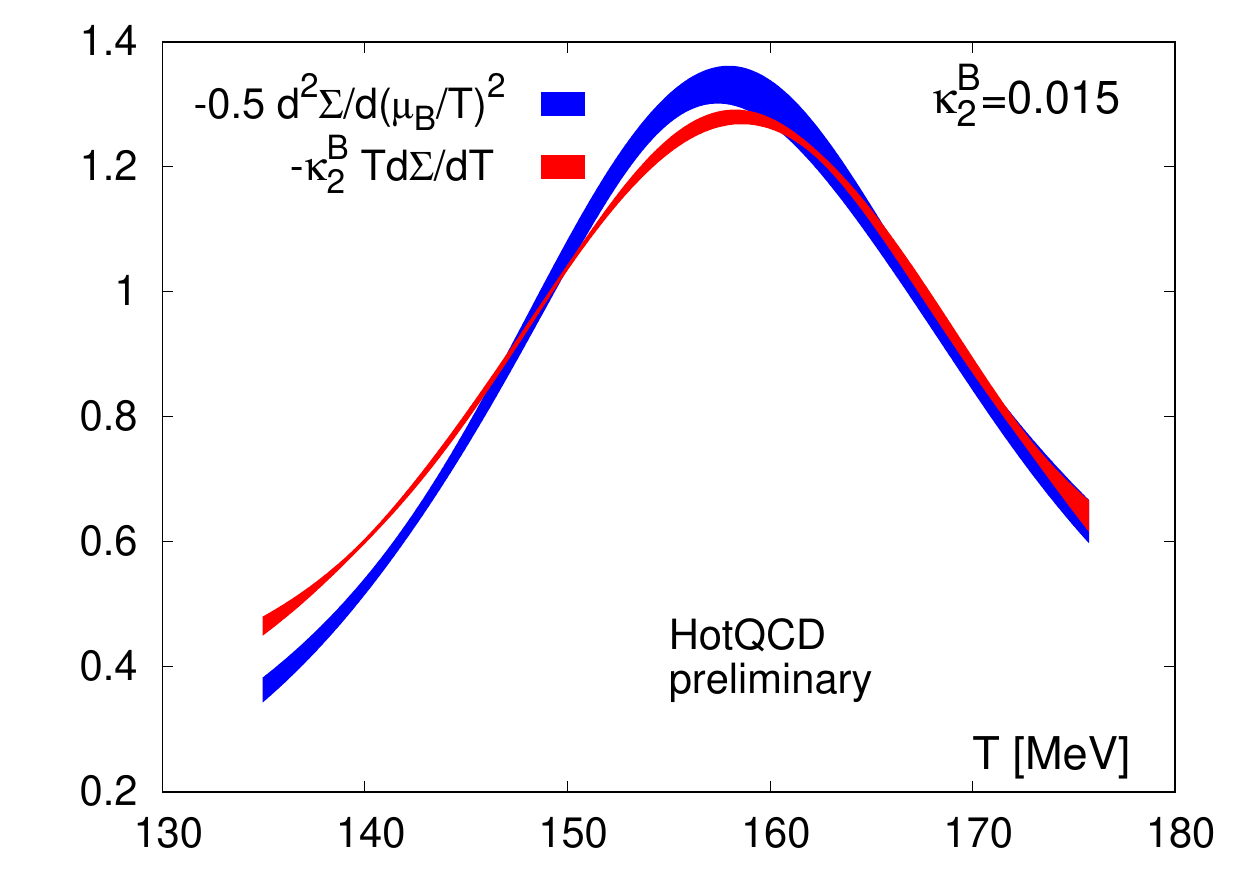}
        \caption{Derivatives of the chiral order parameter with
        respect to temperature and baryon chemical potentials, respectively.
        Shown are results for $N_\tau=12$.
}
\label{fig:scaling}
\end{wrapfigure}
\noindent
A test of this relation is shown in Fig.~\ref{fig:scaling},
where $\kappa_2^{B,0}\equiv \kappa_2^{B}$ has been assumed. This indeed
suggests that the curvature of the chiral phase transition line is
similar in magnitude to that of the pseudo-critical line at physical
values of the quark masses.

\vspace{-0.2cm}
\section{Higher order cumulants in the \\
crossover region}

The sketch of the QCD phase diagram shown in Fig.~\ref{fig:phasediagram},
which qualitatively is consistent with model calculations for the 
quark mass dependence of transition lines in the QCD phase diagram
\cite{Halasz:1998qr,Stephanov:2006dn,Buballa:2018hux},
suggests
that a possible critical point at physical values of the quark
masses is located at a temperature $T_{cep}$ below the chiral phase transition
temperature $T_c^0$. If this is correct, it has significant consequences
also for the properties of higher order cumulants of conserved charge 
fluctuations.

Cumulants of conserved charge fluctuations, evaluated at vanishing
chemical potentials ($\mu_{B,Q,S}$), appear as expansion
coefficients in Taylor series for thermodynamic quantities. The relative
magnitude of subsequent expansion coefficients controls the convergence
of these expansions and determines their radius of convergence. The pattern
of sign changes in these expansion coefficients provides information on
the location of singularities in the plane of complex-valued 
chemical potentials which cause the breakdown of the Taylor 
expansions. E.g., for a series of the form $\sum_x c_nx^n$ the 
singularity determining the radius of convergence lies on the real-x axis,
if an $n_0$ exists such that all expansion coefficients $c_n$ are positive 
for all $n>n_0$ \cite{Gaunt} (see also discussion in \cite{Allton:2005gk}). Only
in this case the radius of convergence can be unambiguously related to the
existence of a phase transition in the thermodynamic system under 
consideration. One thus
may examine the sign of subsequent expansion coefficients and their 
relative magnitude in order to judge whether or not the convergence
of a Taylor series is limited by the appearance of a phase transition 
for some real-valued chemical potential.

At small values of the chemical potentials the QCD partition function
may be expanded in a Taylor series. E.g. the pressure can be written as
\begin{equation}
\frac{P}{T^4} = \frac{1}{VT^3}\ln~Z(T,V,\hmu_u,\hmu_d,\hmu_s) = \sum_{i,j,k=0}^\infty%
\frac{\chi_{ijk}^{BQS}}{i!j!\,k!} \hmu_B^i \hmu_Q^j \hmu_S^k \; ,
\label{Pdefinition}
\end{equation}
with $\chi_{000}^{BQS}\equiv P(T,0)/T^4$ and $\hmu_X=\mu_X/T$. The generalized susceptibilities
are given as derivatives of $P/T^4$ at vanishing values of the
conserved charge chemical potentials,
\begin{equation}
\chi_{ijk}^{BQS}\equiv \chi_{ijk}^{BQS}(T) =\left.
\frac{\partial P(T,\hmu)/T^4}{\partial\hmu_B^i \partial\hmu_Q^j \partial\hmu_S^k}\right|_{\hmu=0} \; .
\label{suscept}
\end{equation}

If, at some value of the temperature, the radius of convergence of the Taylor 
series for the pressure arises from a singularity in the complex-$\mu$ plane, 
one should find that Taylor expansion coefficients will have
an irregular sign structure, {\it i.e.} at this temperature positive and 
negative expansion coefficients will appear in the Taylor series. Such 
changes of 
sign are indeed observed for various cumulants of conserved charge 
fluctuations, starting with sixth order expansion coefficients. 
Although not rigorous in the mathematical sense stated above, these 
sign changes suggest that Taylor expansions in this temperature range are
not limited by a physical singularity related to a phase transition, but
by some singularity in the complex-$\mu$ plane.

\begin{wrapfigure}{l}{0.5\textwidth}
        \includegraphics[width=0.45\textwidth]{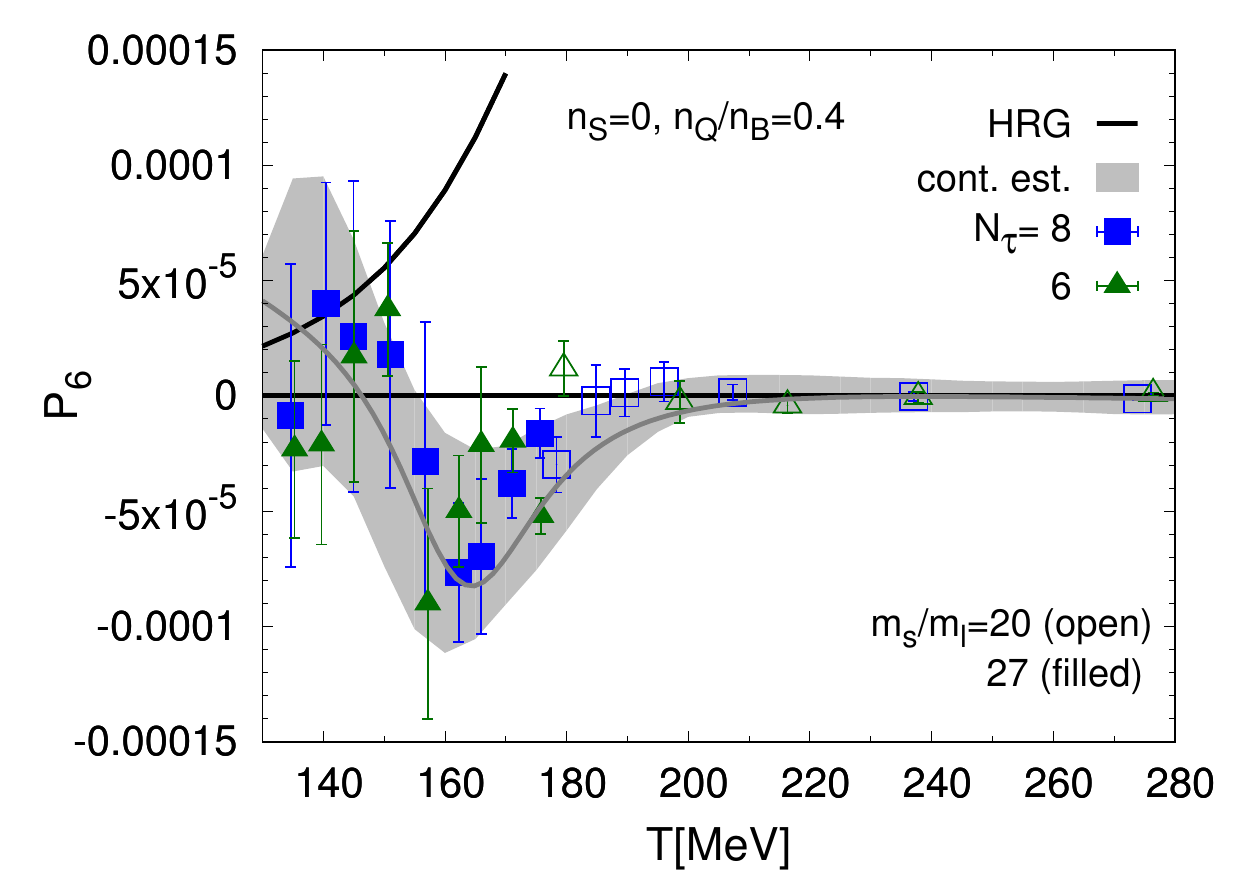}
        \caption{Temperature dependence of the sixth order
        expansion coefficient of the pressure in (2+1)-flavor QCD at
        vanishing net strangeness and fixed electric charge to baryon number
	density, $n_Q/n_B=0.4$ \cite{Bazavov:2017dus}.
}
\label{fig:P6}
\end{wrapfigure}
\noindent
In Fig.~\ref{fig:P6} we show the sixth order expansion coefficient
of the pressure for the case of vanishing net strangeness and a
fixed electric charge to baryon-number density, $n_Q/n_B=0.4$ 
\cite{Bazavov:2017dus},
\begin{eqnarray}
	\frac{P}{T^4} &=& P_0+ P_2 \hat{\mu}_B^2 +
	P_4 \hat{\mu}_B^4 +
	P_6 \hat{\mu}_B^6 \nonumber \\
	&&+ {\cal O}(\hat{\mu}_B^8)\; .
	\label{Pn}
\end{eqnarray}
While the expansion coefficients up to ${\cal O}(\mu_B^4)$ are all
positive \cite{Bazavov:2017dus}, the sixth order expansion coefficient, 
$P_6$, starts
to change sign with increasing temperature, {\it i.e. $P_6 < 0$}
for $T\gsim 150$ MeV. These sign changes are expected to become
more frequent and start at lower temperatures in higher orders
of the expansion. 

\begin{figure}[t]
        \centering
        \includegraphics[height=4.9cm]{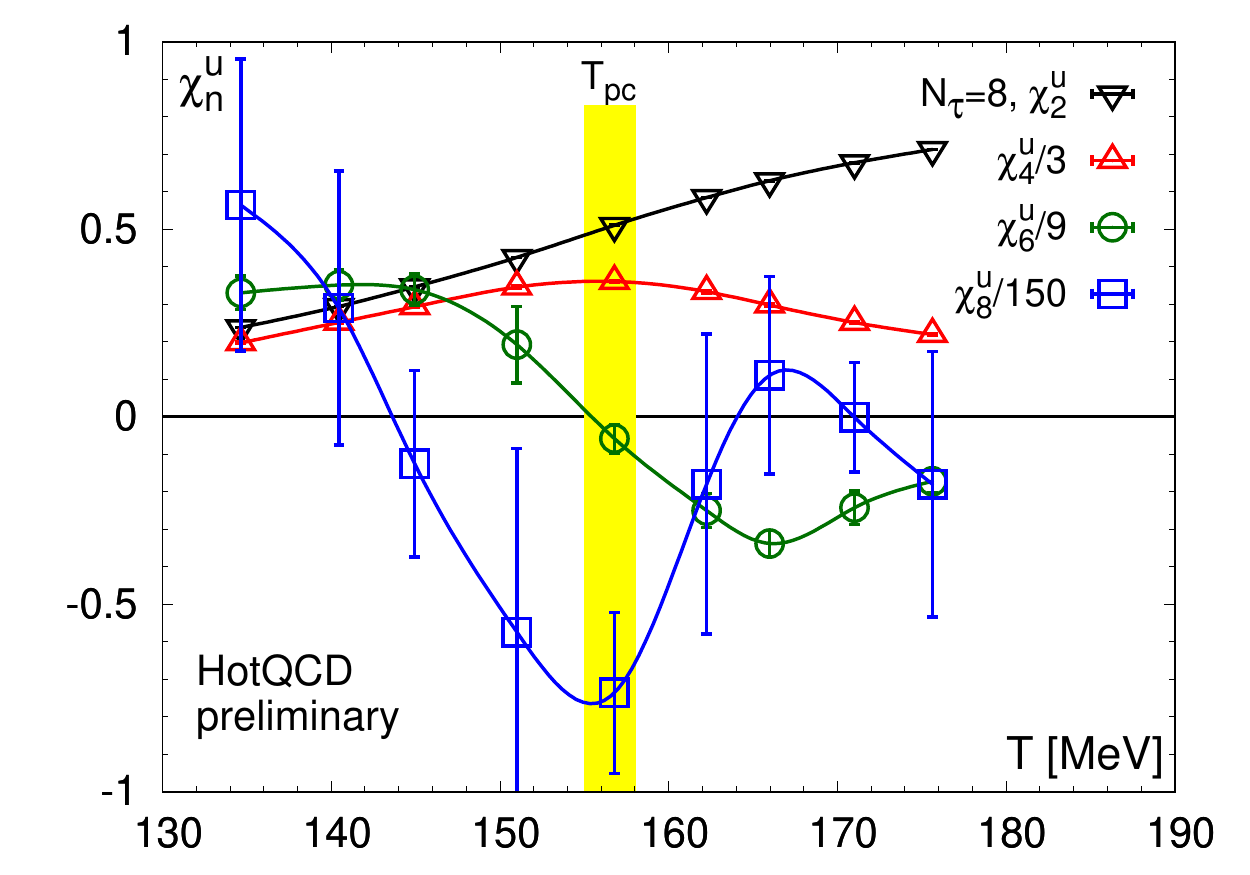}\hspace{0.4cm}
        \includegraphics[height=4.9cm]{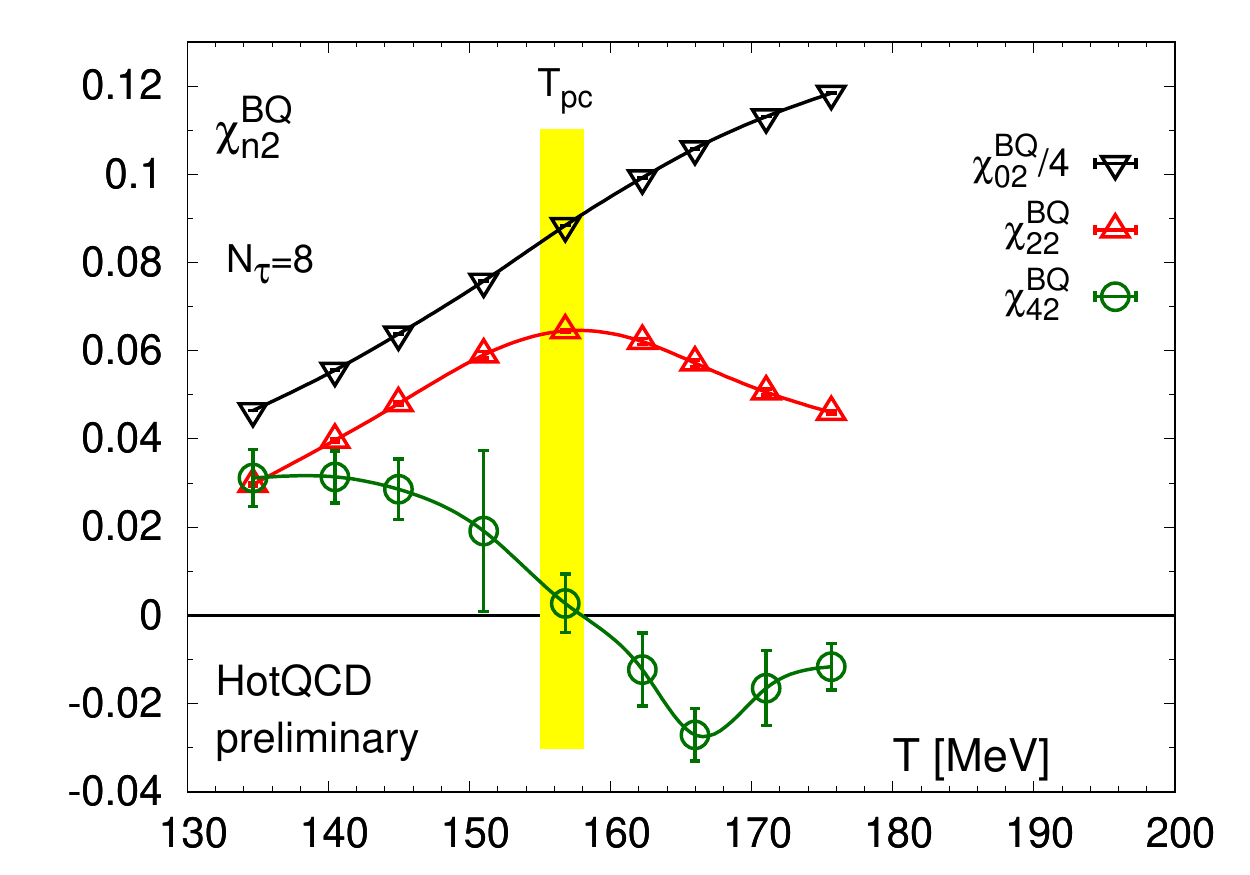}
        \caption{{\it Left:} Temperature dependence of 
	up to eight order cumulants of net up-quark-number fluctuations 
	calculated on lattices with temporal extent $N_\tau =8$
        {\it Right:} Expansion coefficients of net electric charge 
	fluctuations for the case of vanishing electric charge and strangeness 
	chemical potentials. In both figures the lines are smooth spline 
	interpolations drawn to guide the eye.
}
\label{fig:high}
\end{figure}

The irregular sign structure becomes more apparent in simpler cumulants like 
the net up-quark-number cumulants, which
are statistically easier to control. Up to eight order cumulants are shown 
in Fig.~\ref{fig:high}~(left). 
As can be seen, the sign of $\chi_{n+2}^u(T)$ can be 
deduced from the temperature derivative of $\chi_{n}^u(T)$, as suggested
by Eq.~\ref{reduced}. 
Similar behavior is found for the expansion coefficients of the
quadratic net electric charge fluctuations at non-zero 
baryon chemical potential,
\begin{equation}
	\chi_2^Q(T,\mu_B) = \chi_{02}^{BQ}(T) + \frac{1}{2}\chi_{22}^{BQ}(T) \hat{\mu}_B^2
	+\frac{1}{24}\chi_{42}^{BQ}(T) \hat{\mu}_B^4 + {\cal O}(\mu_B^6) \; ,
	\label{chiQ }
\end{equation}
where, for simplicity, we have set $\mu_Q=\mu_S=0$. The first three
expansion coefficients are shown in Fig.~\ref{fig:high}~(right). 
We note that $\chi_{42}^{BQ}$ vanishes at the temperature where 
$\chi_{22}^{BQ}$ has its maximum. Also these expansion coefficients thus
seem to be in accordance with the pattern resulting from Eq.~\ref{reduced}
in the scaling regime, 
{\it i.e.} two derivatives with respect to the baryon chemical potential
are proportional to a single derivative with respect to temperature. This
leads to the expectation that the eight order cumulants, $\chi_{62}^{BQ}$,
will be negative in the temperature range $T\in [135~{\rm MeV}:165~{\rm MeV}]$.
At high temperature subsequent expansion coefficients 
thus show an irregular sign structure, which is in accordance
with the expectation
that for physical quark mass values a possible critical endpoint in the QCD 
phase diagram will be located at a temperature below the chiral phase 
transition temperature $T_c^0$.

\section{Conclusions}
\vspace{-0.2cm}
New results on the chiral phase transition temperature $T_c^0$ in (2+1)-flavor QCD 
suggests that this temperature is well below the pseudo-critical temperature $T_{pc}$
at physical values of the light and strange quark masses. Moreover, it is found that
many $6^{th}$ and higher order cumulants of conserved charge fluctuations are no
longer strictly positive but start showing an irregular sign structure at 
temperatures $T\gsim T_c^0$. This suggests that a possible second order phase
transition at physical values of the quark masses and for non-vanishing
baryon chemical potential can occur only at a temperature $T_{cep}< T_c^0$,
if it exists at all. 

\section*{Acknowledgement}
\vspace{-0.2cm}
This work was supported in part through Contract No. DE-SC001270 with the
U.S. Department of Energy,
the Deutsche Forschungsgemeinschaft (DFG) through the  CRC-TR 211 
"Strong-interaction matter under extreme conditions", grant number 
315477589 - TRR 211,
and the grant 05P18PBCA1 of the German Bundesministerium f\"ur Bildung und 
Forschung.

\vspace{-0.2cm}
\bibliographystyle{JHEP}
\bibliography{karsch}
\end{document}